\documentclass[aps,pra,twocolumn,showpacs,groupedaddress,floatfix,nofootinbib]{revtex4}
\usepackage{graphicx}
\usepackage{amsmath}
\usepackage{epsfig}
\usepackage{helvet}
\usepackage{amssymb}

\newcommand{\tr}{\mbox{tr}}

\def\sh{h^{\diamond}}
\def\so{o^{\diamond}}

\def\sw{w^{\diamond}}
\def\sS{\mathcal{S}^{\diamond}}

\def\sphi{\phi^{\diamond}}
\def\shphi{\hat{\phi}^{\diamond}}
\def\srho{\rho^{\diamond}}
\def\slambda{\lambda^{\diamond}}
\def\sDelta{\Delta^{\diamond}}
\def\I{\mathbb{I}}
\def\sI{\mathbb{I}^{\diamond}}
\def\HI{H_{\mbox{\tiny Ising}}}

\def\sE{E^{\diamond}}

\begin{document}

\title{Boundary quantum critical phenomena with entanglement renormalization}

\author{G. Evenbly$^{1}$}
\author{R. N. C. Pfeifer$^{1}$}
\author{V. Pic\'o$^{2}$}
\author{S. Iblisdir$^{2}$}
\author{L. Tagliacozzo$^{1}$}
\author{I. P. McCulloch$^{1}$}
\author{G. Vidal$^{1}$}
\affiliation{$^{1}$School of Mathematics and Physics, the University of Queensland, Brisbane 4072, Australia}
\affiliation{$^{2}$Dpt. Estructura i Constituents de la Materia, Universitat Barcelona, 08028 Barcelona, Spain}

\date{\today}

\begin{abstract}
We extend the formalism of entanglement renormalization to the study of boundary critical phenomena. The multi-scale entanglement renormalization ansatz (MERA), in its scale invariant version, offers a very compact approximation to quantum critical ground states. Here we show that, by adding a boundary to the scale invariant MERA, an accurate approximation to the critical ground state of an infinite chain with a boundary is obtained, from which one can extract boundary scaling operators and their scaling dimensions. Our construction, valid for arbitrary critical systems, produces an effective chain with explicit separation of energy scales that relates to Wilson's RG formulation of the Kondo problem. We test the approach by studying the quantum critical Ising model with free and fixed boundary conditions.
\end{abstract}


\maketitle

The multi-scale entanglement renormalization ansatz (MERA) \cite{ER,MERA} is a tensor network introduced to efficiently represent ground states of quantum many-body systems on a lattice. It results naturally from a real space renormalization group (RG) transformation that employs unitary tensors (\emph{disentanglers}) to remove short-range entanglement from the system, a process known as \emph{entanglement renormalization} \cite{ER}. The use of disentanglers is a key difference between the MERA and the matrix product state (MPS) \cite{MPS}, another tensor network ansatz for quantum spin chains that is at the heart of Wilson's numerical RG (NRG) \cite{Wilson} and of White's density matrix RG (DMRG) \cite{DMRG}. Both MPS and MERA can be used to describe ground states of translation invariant systems. In addition, however, and thanks to the disentanglers, the MERA can also incorporate scale invariance, thus becoming a natural ansatz to investigate fixed points of the RG flow. It has indeed been used to study non-critical RG fixed points, corresponding to systems with topological order \cite{Topo}, and critical RG fixed points, corresponding to continuous quantum phase transitions \cite{ER,MERA,Free,Transfer,MERACFT,Fazio}.

The goal of this paper is to extend \emph{entanglement renormalization} to the study of boundary quantum critical phenomena. Given a semi-infinite 1D lattice $\mathcal{L}$ at a (scale invariant) quantum critical point, we propose the use of the scale invariant MERA \emph{with a boundary} as an ansatz for its ground state. The boundary of the MERA consists of a semi-infinite MPS that represents the ground state of an effective lattice $\tilde{\mathcal{L}}$ obtained from $\mathcal{L}$ through an inhomogeneous coarse-graining, and where each site represents a different length scale. As in Wilson's formulation of the Kondo problem \cite{Wilson}, lattice $\tilde{\mathcal{L}}$ exhibits explicit scale separation of energies. This MPS is characterized by a single tensor from which boundary scaling operators and boundary scaling dimension, as well as the boundary contribution to the ground state energy and entanglement entropy can be extracted (see appendices). We demonstrate our approach by studying a semi-infinite critical quantum Ising chain with free and fixed boundary conditions.

\begin{figure}
\begin{center}
\includegraphics[width=8cm]{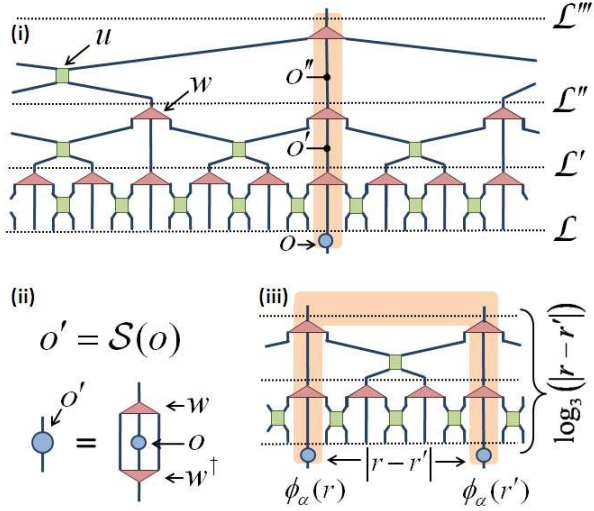}
\caption{(Color online) (i) The bulk (scale invariant) MERA is characterized by a bulk disentangler $u$ and bulk isometry $w$. (ii) Scaling superoperator $\mathcal{S}$ for one-site local operators $o$, in terms of the bulk isometry $w$. (iii) The two-point correlator $\langle \phi_{\alpha}(r) \phi_{\alpha}(r')\rangle$ for selected sites $r$ and $r'$ can be computed by coarse-graining the lattice $\log_3 (|r-r'|)$ times, after which  the two copies of $\phi_{\alpha}$ are nearest neighbors and fuse into the identity with unit amplitude (by normalization of $\phi_{\alpha}$). Each coarse-graning step multiplies $\phi_{\alpha}$ by $3^{-\Delta_{\alpha}}$, resulting in the power-law decay of Eq. \ref{eq:bulk:2point} (see Ref. \cite{MERACFT} for details).} 
\label{fig:bulkMERA}
\end{center}
\end{figure}

\emph{Bulk MERA.---} Recall that the scale invariant MERA for an infinite critical chain \cite{ER,MERA,Free,Transfer,MERACFT,Fazio} is characterized by a unique pair of bulk tensors, namely a disentangler $u$ and an isometry $w$, distributed in layers according to Fig. \ref{fig:bulkMERA}(i). A layer of disentanglers and isometries defines a real space RG transformation that can be used to coarse-grain the original lattice $\mathcal{L}$, producing a sequence of increasingly coarse-grained lattices $\{\mathcal{L}, \mathcal{L}', \mathcal{L}'', \cdots\}$. Under coarse-graining, a local operator $o$ transforms according to the scaling superoperator $\mathcal{S}$, Fig. \ref{fig:bulkMERA}(ii),
\begin{equation}
o  \stackrel{\mathcal{S}}{\longrightarrow}  o' \stackrel{\mathcal{S}}{\longrightarrow} o''  ~\cdots~
\label{eq:bulkSo}
\end{equation}
A \emph{scaling operator} $\phi_{\alpha}$ is a special type of operator that, under coarse-graining, transforms into itself times some scaling factor. The scaling operators $\phi_{\alpha}$ and scaling dimensions $\Delta_{\alpha}$ are obtained from the eigenvalue decomposition of the scaling superoperator $\mathcal{S}$ \cite{Transfer,MERACFT},
\begin{equation}
	\mathcal{S}(\phi_{\alpha}) = \lambda_{\alpha} \phi_{\alpha}, ~~~~~\Delta_{\alpha} \equiv -\log_3 \lambda_{\alpha},	
\label{eq:bulkS}
\end{equation}
where the base $3$ of the logarithm reflects the fact that the coarse-graining transformation maps three sites into one.
Recall that from the scaling dimensions, which govern the decay of two-point correlators in the bulk, Fig. \ref{fig:bulkMERA}(iii),
\begin{equation}
	\langle \phi_{\alpha} (r) \phi_{\alpha} (r')\rangle = \frac{1}{|r-r'|^{2\Delta_{\alpha}}},
\label{eq:bulk:2point}
\end{equation}
one can extract the critical exponents of the model. More generally, the conformal data that characterizes the conformal field theory (CFT) \cite{CFT} associated to the critical chains can be extracted from the scale invariant MERA \cite{MERACFT}. [In this paper we only consider \emph{one-site} scaling operators, placed on special sites in the lattice, as opposed to more general two-site scaling operators, placed on arbitrary sites (see Ref. \cite{MERACFT} for more details), since the latter are not essential to the present discussion on boundary critical phenomena.]

\begin{figure}
\begin{center}
\includegraphics[width=8cm]{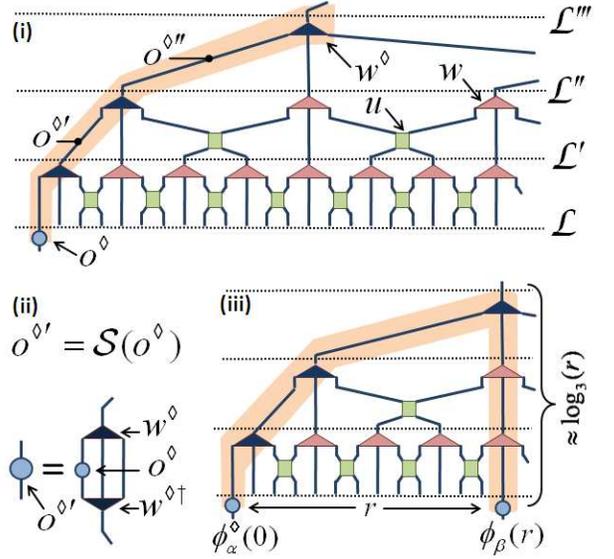}
\caption{(Color online) (i) The boundary (scale invariant) MERA is characterized by the bulk tensors $u$ and $w$ and a boundary isometry $\sw$. Copies of the boundary isometry $\sw$ are connected together. (ii) Scaling superoperator $\sS$ in terms of the boundary isometry $\sw$. (iii) The two-point correlator $\langle \sphi_{\alpha} (0) \phi_{\beta} (r)\rangle$, for $r = (3^{t+1}-1)/2$, is obtained by coarse-graining the lattice $t \approx \log_3 (r)$ times, after which $\sphi_{\alpha}$ and $\phi_{\beta}$ are nearest neighbors and fuse into the identity with amplitude $C_{0\beta}$, see Eq. \ref{eq:BoundaryBulk}. } 
\label{fig:boundMERA}
\end{center}
\end{figure}

\emph{Boundary MERA}.--- In order to represent the ground state of a semi-infinite, quantum critical spin chain, we propose to use the boundary MERA described in Fig. \ref{fig:boundMERA}(i), which is made of semi-infinite layers of copies of the bulk disentangler $u$ and bulk isometry $w$, together with copies of a boundary isometry $\sw$ placed at the ends of the layers. This ansatz defines a real space RG transformation that, away from the boundary, is identical to that of the bulk MERA, Eq. \ref{eq:bulkSo}, but such that a local operator $\so$ at the boundary transforms according to the \emph{boundary scaling superoperator} $\sS$, defined in terms of the boundary isometry $\sw$, Fig. \ref{fig:boundMERA}(ii),
\begin{equation}
\so  \stackrel{\sS}{\longrightarrow}  {\so}' \stackrel{\sS}{\longrightarrow} {\so}''  ~\cdots~
\end{equation}
This allows us to identify a new set of scaling operators, namely the \emph{boundary scaling operators} $\sphi_{\alpha}$, and scaling dimensions $\sDelta_{\alpha}$, which 
are obtained from the eigenvalue decomposition of $\sS$,
\begin{equation}
	\sS(\sphi_{\alpha}) = \slambda_{\alpha} \sphi_{\alpha}, ~~~~~\sDelta_{\alpha} \equiv -\log_3 \slambda_{\alpha}.	
	\label{eq:sS}
\end{equation}
A correlator between a boundary scaling operator $\sphi_{\alpha}$ and a bulk scaling operator $\phi_{\beta}$, Fig. \ref{fig:boundMERA}(iii), reads
\begin{equation}
		\langle \sphi_{\alpha} (0) \phi_{\beta} (r)\rangle \approx \frac{C_{\alpha\beta}}{{r}^{\sDelta_{\alpha} + \Delta_{\beta}}}.
\label{eq:BoundaryBulk}
\end{equation}
A rather significant, particular case of the above is when we choose the boundary scaling operator to be the identity operator $\I$, which has vanishing scaling dimension,
\begin{equation}
		\langle \phi_{\beta} (r)\rangle \approx \frac{C_{0\beta}}{{r}^{\Delta_{\beta}}}.
		\label{eq:newBulk}
\end{equation}
Recall that in a critical system without a boundary, $\langle \phi_{\beta} (r)\rangle_{\mbox{\tiny bulk}} =0$ for any scaling operator (other than the identity). Eq. \ref{eq:newBulk} tells us that in the boundary MERA, the presence of the boundary is felt everywhere in the bulk, with an intensity that decays as a power law with the distance to the boundary. This is precisely one of the trademarks of boundary critical systems, as described by a boundary CFT (BCFT) \cite{CFT,CardyBCFT}.

\begin{figure}
\begin{center}
\includegraphics[width=8cm]{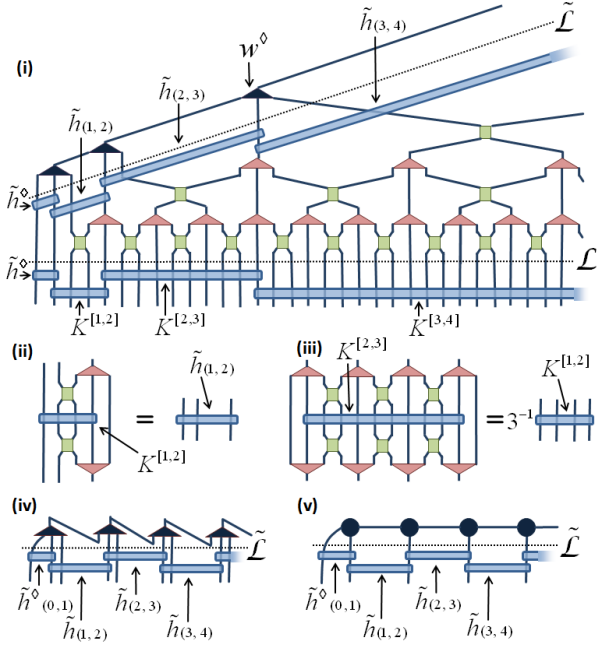}
\caption{(Color online) (i) The effective lattice $\tilde{\mathcal{L}}$ is obtained by coarse-graining $\mathcal{L}$ in an inhomogeneous way. Hamiltonian term $K^{[t,t+1]}$ involving $3^{t}$ sites in $\mathcal{L}$ becomes the term $3^{1-t} \tilde{h}(t,t+1)$ on $\tilde{\mathcal{L}}$. (ii) Definition of $\tilde{h}^{[1,2]}$ in terms of $K^{[1,2]}$, $u$ and $w$. (iii) Applying one layer of coarse-graining to $K^{[2,3]}$ produces $3^{-1}K^{[1,2]}$. (iv) Copies of $\sw$ are used to describe the ground state of $\tilde{H}$ in $\mathcal{\tilde L}$, Eq. \ref{eq:HH}. (v) The ansatz is equivalent to a MPS.} 
\label{fig:boundLattb}
\end{center}
\end{figure}

\emph{Effective lattice}.--- 
Let
\begin{equation}
	H = {\sh}(0) + \sum_{r=0}^{\infty} h(r,r+1)
	\label{eq:H}
\end{equation}
be the Hamiltonian of the semi-infinite chain $\mathcal{L}$, where $\sh$ is a boundary term and $h(r,r+1)\equiv h$ is a (constant) two-site interaction term corresponding to a critical fixed-point bulk Hamiltonian free of irrelevant operators and with vanishing ground state energy. That is, we assume that $h$ is a (\emph{two-site} \cite{MERACFT}) scaling operator with scaling dimension $\Delta=2$ (but see remark on transitional layers below), and thus normalized to $\langle h \rangle_{\mbox{\tiny bulk}}=0$. Let 
\begin{equation}
	K^{[t,t+1]} \equiv \sum_{r=r_{t}}^{r_{t+1}-1} h(r,r+1), ~~~~~r_{t} \equiv (3^{t}-1)/2,
\end{equation}
be an operator that collects the $3^{t}$ two-site terms $h$ included in the interval of sites $[r_t,r_{t+1}]$ of $\mathcal{L}$. The boundary MERA defines an effective lattice $\tilde{\mathcal{L}}$, with sites labelled by $t\in \{0,1,2,\cdots \infty\}$, that results from coarse-graining the original lattice $\mathcal{L}$ in an inhomogeneous way, such that site $t \in \tilde{\mathcal{L}}$ corresponds to $O(3^{t})$ sites of $\mathcal{L}$, see Fig. \ref{fig:boundLattb}(i).  Under the inhomogeneous coarse-graining, the original Hamiltonian becomes
\begin{equation}
	\tilde{H} = \tilde{h}^{\diamond}(0,1) + \sum_{t=1}^{\infty} \Lambda^{1 - t} ~\tilde{h}(t,t+1),
	\label{eq:HH}
\end{equation}
where $\tilde{h}^{\diamond}(0,1)$ corresponds to $\sh(0) + h(0,1)$ in Eq. \ref{eq:H}, $\Lambda$ is just the scaling factor ($\Lambda = 3$), and the two-site term $\Lambda^{1-t}~\tilde{h}(t,t+1)$ results from coarse-graining $K^{[t,t+1]}$. For instance, $\tilde{h}(1,2)$ comes from coarse-graining $K^{[1,2]} \equiv h(1,2)+h(2,3)+h(3,4)$ as in Fig. \ref{fig:boundLattb}(ii); the term $3^{-1} \tilde{h}(2,3)$ comes from coarse-graining $K^{[2,3]}$ into $3^{-1} K^{[1,2]}$, Fig. \ref{fig:boundLattb}(iii), and then $K^{[1,2]}$ into $\tilde{h}$ as before; more generally, $K^{[t,t+1]}$ is first coarse-grained into $3^{1-t} K^{[1,2]}$ and then $K^{[1,2]}$ again into $\tilde{h}$. To understand the origin of the scaling factor $3^{-t}$, we notice that under coarse-graining in the bulk, $K^{[t,t+1]}$ (made of $3^{t}$ terms $h$) becomes proportional to $K^{[t-1,t]}$ (made of $3^{t-1}$ terms $h$)
\begin{equation}
	K^{[t,t+1]}~~n\stackrel{\mbox{\tiny RG}}{\longrightarrow}~~3^{-1} K^{[t-1,t]},~~~~~(t>1)
\end{equation}
where the factor $3^{-1} = 3^{-2}\times 3$ is due to the scaling dimension $\Delta=2$ of $h$ (factor $3^{-\Delta} = 3^{-2}$) and the fact that each term $h$ in $K^{[t-1,t]}$ comes from three terms $h$ in $K^{[t,t+1]}$ (factor $3$). Then, after $t-1$ iterations of the RG transformation, $K^{[t,t+1]}$ indeed becomes
\begin{equation}
	K^{[t,t+1]} ~~\stackrel{\mbox{\tiny RG}}{\longrightarrow} ~~3^{1-t} K^{[1,2]},~~~~~(t\geq 1).
\end{equation}

The effective Hamiltonian $\tilde{H}$ is of the form derived by Wilson as part of his resolution of the Kondo problem \cite{Wilson}. It describes a semi-infinite chain of sites that interact with nearest neighbors with the same interaction term $\tilde{h}$, which is multiplied by a factor $\Lambda^{1-t}$ decreasing exponentially fast with the distance to the boundary. Notice, however, that while Wilson's derivation was for free fermions and the limit $\Lambda \rightarrow 1$ was eventually taken, here we started with a generic critical Hamiltonian and $\Lambda$ remains fixed at $\Lambda=3$. In a recent paper \cite{Okunishi}, Okunishi proposed an interesting generalization of Wilson's NRG approach that also considered a generic critical Hamiltonian, although the scale $\Lambda>1$ in Eq. \ref{eq:HH} was simply introduced `\emph{by hand}' as a regulator of the gapless spectrum and made very close to 1 (e.g. $\Lambda = 1.02$) to avoid an `undesired perturbation' in its degeneracy structure, and $\tilde{h}$ was chosen to be simply equal to $h$.

\begin{figure}
\begin{center}
\includegraphics[width=8cm]{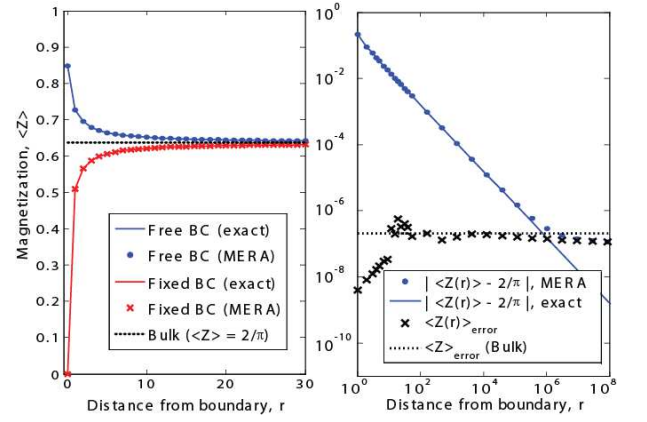}
\caption{(Color online) Left: numerical estimate $\langle Z(r) \rangle$ for free and fixed BC obtained with a boundary MERA. The exact solution approaches the bulk value $2/\pi$ as $\approx r^{-1}$. Right: the error in $\langle Z(r) \rangle$ for free BC (very similar to that for fixed BC) shows that the non-vanishing expectation value of the bulk scaling operators, Eq. \ref{eq:newBulk}, is still accurately reproduced thousands of sites away from the boundary.} 
\label{fig:Zmag}
\end{center}
\end{figure}

\emph{Optimization}.--- Given a critical Hamiltonian $H$, Eq. \ref{eq:H}, the bulk tensors $u$ and $w$ are computed using the optimization algorithm for the bulk scale invariant MERA discussed in Refs. \cite{MERACFT,MERAalgorithm}. Then $H$ is coarse-grained into the effective Hamiltonian $\tilde{H}$ of Eq. \ref{eq:HH}. The boundary isometry $\sw$ is obtained with a simplified version (replacing the MERA with an MPS) of the energy minimization techniques used in Refs. \cite{MERACFT,MERAalgorithm} for the scale invariant MERA. An important point is that, both for the bulk and boundary MERA, in practical simulations we consider a few (e.g. three in the example below) transitional layers made of tensors $\{u_1,w_1,\sw_1\}$, $\{u_2,w_2,\sw_2\}$, etc. that are different from the fixed-point tensors $\{u,w,\sw\}$. As discussed in Ref. \cite{MERACFT,MERAalgorithm}, these transitional layers allow us to modify the vector space dimension $d$ of one site of the original model to some larger value $\chi$ and, if (contrary to our assumption above) $H$ contains irrelevant operators, significantly diminish their effect.

\begin{figure}
\begin{center}
\includegraphics[width=8.5cm]{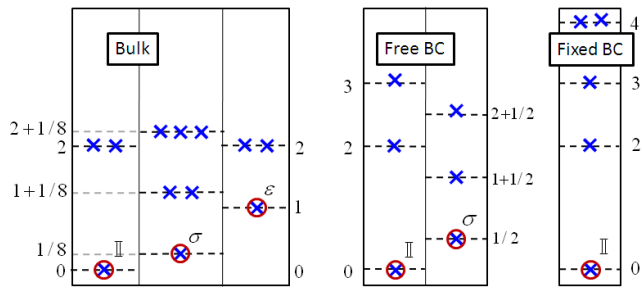}
\caption{(Color online) A few boundary scaling dimensions, organized in conformal towers \cite{CFT}, for the quantum Ising model with free and fixed BC. We include bulk scaling dimensions for comparison. The primary fields identity $\mathbb{I}$, spin $\sigma$ and energy $\epsilon$ are identified with a circle. The boundary MERA accurately reproduces the smallest scaling dimensions of each conformal tower.} 
\label{fig:ScaleDim}
\end{center}
\end{figure}

\emph{Example}.--- In order to test the performance of the approach, we use a boundary MERA \cite{chi} to approximate the ground state of the critical quantum Ising model on a semi-infinite chain,
\begin{equation}
	\HI = \eta X(0) -  \sum_{r=0}^{\infty} \left( X(r)X(r+1) - Z(r) \right)
\end{equation}
where $X$ and $Z$ are the Pauli matrices and the constant $\eta$ determines whether the system has free BC ($\eta=0$) or fixed BC ($\eta = \pm 1$). 
Fig. \ref{fig:Zmag} shows, through accurate estimates of the expectation value $\langle Z(r) \rangle$ for both free and fixed BC, that the boundary MERA indeed offers a good approximation to the ground state of the original lattice $\mathcal{L}$, in spite of the fact that the ansatz uses, arbitrarily close to the boundary, bulk tensors $u$ and $w$ that have been optimized in the absence of a boundary! Table \ref{table:scaling} shows some boundary scaling dimensions obtained by diagonalizing the boundary scaling superoperator $\sS$. Finally, in Fig. \ref{fig:ScaleDim}, these scaling dimensions are organized according to the conformal towers of the primary fields predicted by BCFT \cite{CFT}. 

\begin{table} [!htbp]
\begin{tabular}{||l|l|l||l|l|l||}
  \hline
  $~~~~~\Delta_{\mbox{\tiny free}}^{\mbox{\tiny BCFT}}$  & $~\Delta^{\mbox{\tiny MERA}}_{\mbox{\tiny $\chi=16$}}~$ & error &

  $~~~\Delta_{\mbox{\tiny fixed}}^{\mbox{\tiny BCFT}}$ & $~\Delta^{\mbox{\tiny MERA}}_{\mbox{\tiny $\chi=16$}}~$ & error\\ \hline

   $(\mathbb{I})~$0 &  $~$0     & -- & $(\mathbb{I})~$0  &  $~$ 0      & --\\

   $(\sigma)$ 0.5 &  $~$0.499  & 0.2$\%$  & $~~~\,$ 2  &  $~$1.992   & 0.4$\%$\\

   $~~~\,$ 1.5 $~$  &  $~$1.503  & 0.18$\%$ & $~~~\,$ 3  &  $~$2.998   & 0.07$\%$\\

   $~~~\,$ 2   $~$  &  $~$2.001  & 0.07$\%$ & $~~~\,$ 4  &  $~$4.005   & 0.12$\%$\\

   $~~~\,$ 2.5 $~$  &  $~$2.553   & 2.1$\%$ & $~~~\,$ 4  &  $~$4.062   & 1.5$\%$\\
  \hline
  \end{tabular} \nonumber
  \caption{Some scaling dimensions for free and fixed BC.}
  \label{table:scaling}
\end{table}

In summary, we have extended the formalism of entanglement renormalization to the study of a critical infinite lattice with a boundary. The boundary (scale-invariant) MERA only depends on two tensors $u$ and $w$ that encode bulk properties and a third tensor $\sw$ that encodes boundary properties. We have seen that this simple ansatz can accurately reproduce the expectation value of local observables near and away from the boundary, as well as the lower part of the spectrum of boundary scaling dimensions of the model. 
By considering a different boundary isometry at each level of coarse-graining, it is also possible to study boundary RG flows (e.g. from free BC to fixed BC). 
Our results provide a numerical route to the study of boundary conformal field theory \cite{CardyBCFT} that may find applications in several areas ranging from condensed matter physics (boundary critical behaviours and quantum impurity problems) to string theory (open strings and D-branes).

Support from the Australian Research Council (APA, FF0668731, DP0878830) and the Spanish Ministerio de Ciencia e Innovaci\'on (RYC-2009-04318) is acknowledged.


\appendix 
\section{Boundary-bulk fusion amplitudes}

In this appendix we explain how to extract the coefficient $C_{\alpha\beta}$ for the boundary-bulk correlator
\begin{equation}
		\langle \sphi_{\alpha} (0) \phi_{\beta} (r)\rangle \approx \frac{C_{\alpha\beta}}{{r}^{\sDelta_{\alpha} + \Delta_{\beta}}},
\label{eq:x}
\end{equation}
and more generally the coefficient $C_{\alpha \beta}^{\gamma}$ that governs the fusion of a boundary scaling operator $\sphi_{\alpha}$ and a bulk scaling operator $\phi_{\beta}$ sitting next to the boundary into another boundary scaling operator $\sphi_{\gamma}$,
\begin{equation}
	\sphi_{\alpha}(0) \times \phi_{\beta}(1) ~~\stackrel{\mbox{\tiny RG}}{\longrightarrow} ~~ \sum_{\gamma} C_{\alpha\beta}^{\gamma} \sphi_{\gamma}(0)
\end{equation} 

\begin{figure}[!htbp]
\begin{center}
\includegraphics[width=8cm]{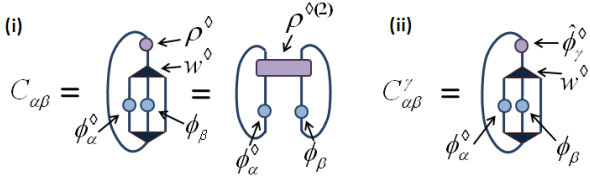}
\caption{(i) Fusion amplitude $C_{\alpha\beta}$. (ii) Fusion amplitude $C_{\alpha\beta}^{\gamma}$.}
\label{fig:OPEcoeff} 
\end{center}
\end{figure}

Recall that the boundary scaling operators $\sphi_{\alpha}$ are obtained from the eigenvalue decomposition of the boundary scaling superoperator $\sS$, Eq. \ref{eq:sS}. We can now consider the set of eigenoperators $\shphi_{\alpha}$ of the superoperator ${\sS}^{*}$ dual to the scaling superoperator $\sS$,
\begin{equation}
	{\sS}^{*}(\shphi_{\alpha}) = \slambda_{\alpha} \shphi_{\alpha},~~~~~~~~
\end{equation}
where $-\log_3 \slambda_{\alpha}$ corresponds to the scaling dimension $\sDelta_{\alpha}$. By construction, the $\shphi_{\alpha}$'s form a biorthonormal set with the $\phi_{\alpha}$'s,
\begin{equation}
	\tr(\sphi_{\alpha}\shphi_{\beta}) = \delta_{\alpha\beta}.
\end{equation}
The boundary identity operator, $\sphi_{0} =\sI$, has scaling dimension $\Delta_0 = 0$ and its counterpart $\shphi_{0}$ is the one-site density matrix $\srho$ of the boundary. From $\srho$ we can compute ${\srho}^{(2)}$ for the two boundary sites, and then evaluate $C_{\alpha\beta}$ as 
\begin{equation}
	C_{\alpha\beta} = \tr \Big[ {\srho}^{(2)}(0,1)\Big(\sphi_{\alpha}(0)\otimes \phi_{\beta}(1)\Big) \Big].
\end{equation}
Fig. \ref{fig:OPEcoeff} shows this expression as well as the one corresponding to $C_{\alpha\beta}^{\gamma}$, which involves $\sphi_{\alpha}$, $\phi_{\beta}$, $\shphi_{\gamma}$ and the boundary isometry $\sw$.

\section{Boundary ground state energy}

\begin{figure}
\begin{center}
\includegraphics[width=8.5cm]{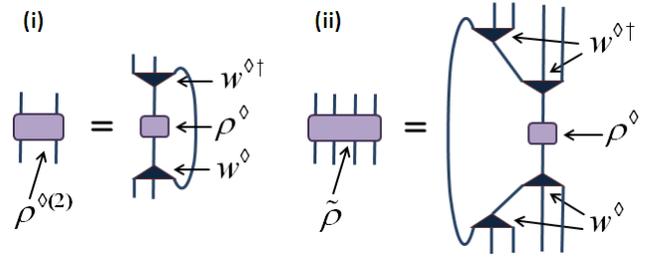}
\caption{(Color online) (i) Density matrix ${\srho}^{(2)}$ in terms of $\srho$ and $\sw$. (ii) Density matrix $\tilde{\rho}$ in terms of $\srho$ and $\sw$. The fixed-point density matrix $\srho$ has been defined in appendix A.} 
\label{fig:boundEnergy}
\end{center}
\end{figure}

In this appendix we explain how to determine the boundary contribution $\sE$ to the ground state energy, defined as the difference between the energy $\langle H \rangle$ of the semi-infinite chain and one half of the ground state energy for an infinite chain,
\begin{equation}
	\sE \equiv \langle H \rangle - \frac{1}{2} \sum_{r=-\infty}^{\infty} \langle h(r,r+1) \rangle_{\mbox{\tiny bulk}}.
\end{equation}
Since in this paper $h$  is normalized such that it vanishes in the bulk (if this was not the case, we would simply choose $h - \langle h\rangle_{\mbox{\tiny bulk}}$ as the new $h$), $\sE$ is the expectation value $\langle H \rangle$ of $H$ in Eq. \ref{eq:H}. In order to evaluate $\langle H \rangle$, we can coarse-grain the lattice $\mathcal{L}$ into the effective lattice $\tilde{\mathcal{L}}$ and use that $\langle H \rangle_{\mathcal{L}} = \langle \tilde{H} \rangle_{\tilde{\mathcal{L}}}$ to write
\begin{equation}
	\sE = \langle \tilde{h}^{\diamond}(0,1) \rangle_{\tilde{\mathcal{L}}} + \sum_{t=1}^{\infty} 3^{1-t} \langle \tilde{h}(t,t+1) \rangle_{\tilde{\mathcal{L}}}.
\end{equation}
For any $t\geq 1$, $\langle \tilde{h}(t,t+1) \rangle_{\tilde{\mathcal{L}}} = \tr(\tilde{\rho} \tilde{h})$, where $\tilde{\rho}$ is the two-site density matrix in $\tilde{\mathcal{L}}$. Recall that, by construction, the semi-infinite MPS describing $\tilde{\mathcal{L}}$ is translation invariant for $t=1,2,\cdots$, so that any two sites $[t,t+1]$ (with $t\geq 1$) are indeed described by the same reduced density matrix $\tilde{\rho}$, which is obtained from the density matrix $\srho$ as indicated in Fig. \ref{fig:boundEnergy}. Since $\sum_{t=1}^{\infty} 3^{1-t} = 3/2$, we arrive at
\begin{equation}
	\sE = \tr({\srho}^{(2)} \tilde{h}^{\diamond}) + \frac{3}{2} \tr(\tilde{\rho}\tilde{h}),
\end{equation}
which is manifestly finite and can be easily computed.

For the quantum Ising model on a semi-infinite lattice we obtain, for free BC, a value $\sE = 0.18169023$ which is remarkably close to the exact solution $\sE_{\mbox{\tiny exact}}= (1/2 -1/\pi) = 0.181690113...$ \cite{CFTHenkel}. For fixed boundary conditions we obtain a value $\sE = -0.45492968$. Based upon the exact solution for finite chains of over a thousand sites, we estimate the error in this value is order $10^{-6}$.  

\begin{figure}[!htbp]
\begin{center}
\includegraphics[width=8cm]{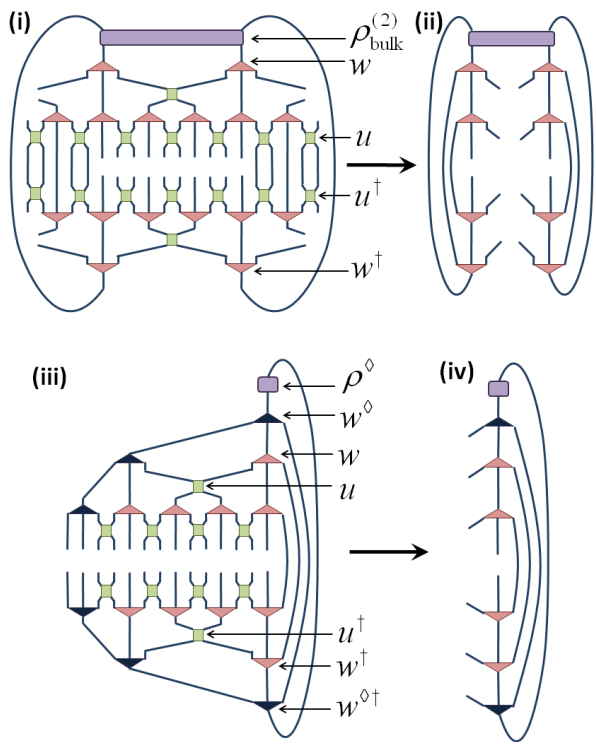}
\caption{(i) Density matrix $\rho$ of a block of $L=3^T+1$ sites in an infinite chain (without boundaries), for $T=2$. The bulk density matrix $\rho^{(2)}_{\mbox{\tiny bulk}}$ is the fixed-point, two-site density matrix of Ref. \cite{MERACFT}.(ii) An effective density matrix with the same spectrum, obtained by removing disentanglers and isometries from $\rho$ (which do not change the spectrum). (iii) Density matrix of a block of $L=(3^T+1)/2$ sites that includes the boundary of a semi-infinite chain, for $T=2$. The fixed-point density matrix $\srho$ has been defined in appendix A. (iv) An effective density matrix with the same spectrum.}\label{fig:boundEnt} 
\end{center}
\end{figure}

\begin{figure}[!htbp]
\begin{center}
\includegraphics[width=5cm]{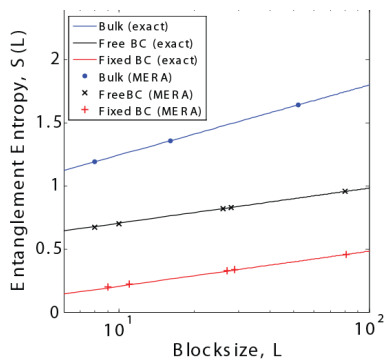}
\caption{Entanglement entropy scaling of bulk, and of free and fixed BC. Continuous lines represent the exact scaling.}\label{fig:boundEntPlot} 
\end{center}
\end{figure}

\section{Boundary entanglement entropy}

In this section we discuss how to compute the boundary entropy from the boundary scale invariant MERA, from which we can extract Affleck-Ludwig's `\emph{ground-state degeneracy}' $g$ associated to the boundary \cite{g}. In the bulk, the von Neumann entropy of a block of $L$ sites scales as \cite{Holzhey,Entropy,Calabrese,Zhou,EntropyB}
\begin{equation}
	S_{\mbox{\tiny bulk}}(L) = \frac{c}{3}\log_2 (L) + c',
\end{equation}
where $c$ is the central charge of the relevant CFT and $c'$ is some non-universal constant,
whereas for a block of $L$ sites at the boundary of a semi-infinite chain, the von Neumann entropy scales as \cite{Calabrese,Zhou,EntropyB}
\begin{equation}
	S_{\mbox{\tiny boundary}}(L) = \frac{c}{6}\log_2 (2L) + \frac{c'}{2} + S_b,
\end{equation}
where $S_b \equiv \log_2 (g)$ is the boundary contribution to the entropy \cite{g}, which depends on the boundary condition. As discussed in \cite{Prep}, the entropy of a region of $L = 3^{T}+1$ sites in an infinite chain is the same as the entropy of a simpler, effective density matrix described in Fig. \ref{fig:boundEnt}(i)-(ii). Similarly, the entropy of a block of $L = (3^{T}+1)/2$ sites including the boundary of a semi-infinite chain is the same as the entropy of a simpler, effective density matrix described in Fig. \ref{fig:boundEnt}(iii)-(iv). Fig. \ref{fig:boundEntPlot} shows the entropies $S_{\mbox{\tiny bulk}}$ and $S_{\mbox{\tiny boundary}}$ for the Ising model. Noticing that $S_b = S_{\mbox{\tiny boundary}}(L) - S_{\mbox{\tiny bulk}}(2L)/2$, from which we can extract $S_b = 0.0007 \approx 0$ (that is, $g=1$) for free BC and $S_b = -0.4992 \approx -1/2$ (or $g=2^{-1/2}$) for fixed BC.

\section{Finite system with open boundary conditions}

In this appendix we discuss how to study the low energy spectrum of a critical Hamiltonian on a finite system with two open boundaries. 
For this purpose, we consider a finite lattice $\mathcal{L}$ made of $N$ sites with a Hamiltonian 
\begin{equation}
	H_{LR} = \sh_L(0) + \sum_{r=0}^{N-2}  h(r,r+1) +  \sh_R(N-1), 
\end{equation}
where $h$ is a critical, fixed-point Hamiltonian term and the left and right open BC are specified by the boundary Hamiltonian terms $\sh_L$ and $\sh_R$. 

\begin{figure}
\begin{center}
\includegraphics[width=8.5cm]{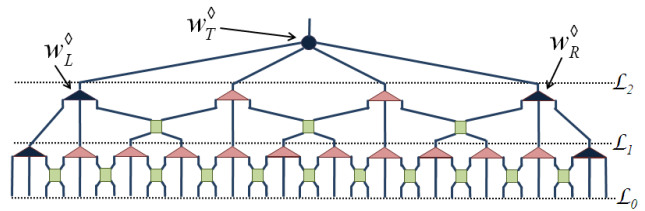}
\caption{(Color online) MERA for a finite system with open BC and critical bulk Hamiltonian, made of $N = 4\times 3^{T}$ sites, $T=2$. Most of this finite-size MERA is filled with copies of the bulk disentangler $u$ and bulk isometry $w$ optimized in an infinite system. The boundary isometries $\sw_L$ and $\sw_R$ are optimized in a semi-infinite system. The top isometry $w_{T}$ is obtained by diagonalizing the coarse-grained Hamiltonian on the effective lattice $\mathcal{L}_{T}$ ($\mathcal{L}_{2}$ in this case).} 
\label{fig:finiteMERA}
\end{center}
\end{figure}

For a system with $N=4\times 3^{T}$ sites, we attempt to describe the low energy subspace of $H_{LR}$ with the MERA described in Fig. \ref{fig:finiteMERA}, which consists of $T$ layers of disentanglers and isometries. Each layer is filled with bulk tensors $u$ and $w$, obtained by analysing an infinite system using the scale invariant MERA algorithm of Ref. \cite{MERACFT}. In adddition, each layer has a boundary isometry $\sw_{L}$ at its left end and a boundary isometry $\sw_{R}$ at its right end, obtained by analysing a semi-infinite system using the boundary scale invariant MERA described in this paper. Finally, the ansatz is completed with a top isometry $w_{T}$, which has an open index labelling (an approximation to) different low energy eigenstates of $H_{LR}$.

\begin{figure}[!htbp]
\begin{center}
\includegraphics[width=6cm]{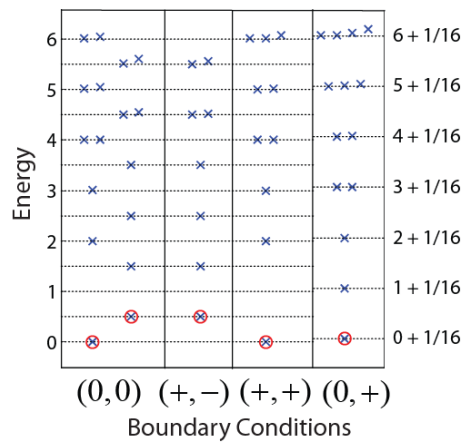}
\caption{Excitation spectra of the quantum Ising model on a finite lattice $\mathcal{L}$ of $N=4 \times 3^5 = 972$ sites (where each site is made of four spins, so that the system has $4\times 3^5 \times 4 = 3888$ spins) with open boundary conditions. The energy is expressed in units such that the gap between descendants is a multiple of unity. All inequivalent combinations of open BC are considered. The different open BC are $(0) = \textrm{free}, (+) = \textrm{fixed(up)}, (-) = \textrm{fixed(down)}$.}\label{fig:FiniteSpect} 
\end{center}
\end{figure}

Thus, the low energy subsapce of $H_{LR}$ is expressed in terms of five tensors $\{u,w, \sw_{L}, \sw_{R}, w_{T}\}$. The bulk disentangler $u$ and bulk isometry $w$ encode the critical properties of the bulk; the boundary isometries $\sw_{L}$ and $\sw_{R}$ encode the critical properties of each of the boundaries separately, and the top isometry $w_{T}$ is the only tensor that contains information about both boundary conditions, as well as about the finite size $N$ of the system. This top isometry $w_{T}$ is simply determined by diagonalizing the effective Hamiltonian $\tilde{H}$ that results from coarse-graining the system with the $T$ layers of disentanglers and isometries.

Fig. \ref{fig:FiniteSpect} shows the low energy spectrum of $H_{LR}$ for all possible combinations of boundary conditions of the critical Ising model. In all cases, they agree with the predictions of BCFT \cite{CardyBCFT}.

\section{Defects and interfaces}

Entanglement renormalization can also be used to study a defect in the bulk of a critical system on an infinite lattice $\mathcal{L}$.

\begin{figure}[!htbp]
\begin{center}
\includegraphics[width=8cm]{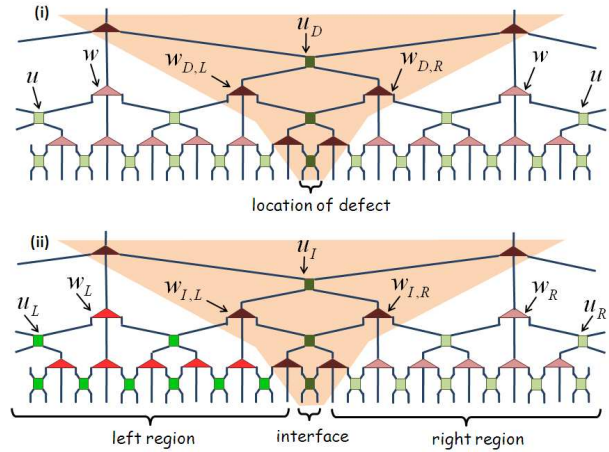}
\caption{(i) Scale invariant MERA for a system with a defect. (ii) Scale invariant MERA to study the interface between two semi-infinite systems at criticality}\label{fig:defectMERA} 
\end{center}
\end{figure}

For this situation we propose a scale invariant MERA with a special stripe of defect disentanglers and isometries, see Fig. \ref{fig:defectMERA}(i). For a fixed-point (i.e. scale invariant) defect, only three defect tensors are needed, namely a defect disentangler $u_{D}$ and left and right defect isometries $w_{D,L}$ and $w_{D,R}$. The scale-invariant MERA with a fixed-point defect is thus characterized by: (i) a pair of bulk tensors $u$ and $v$, which are optimized in the absence of the defect and contain only information about the critical theory in the bulk; and (ii) three defect tensors $u_{D}$, $w_{D,L}$ and $w_{D,R}$, which are optimized by regarding them as a MPS-like structure representing the ground state of a coarse-grained lattice $\tilde{\mathcal{L}}$, similarly as we did in the case of an open boundary, and which contain all the information about the defect, including defect scaling operators and their scaling dimensions.

When tested with the quantum Ising model, we find \cite{Prep} that the scale-invariant MERA with a fixed-point defect is capable of describing all conformal defects, namely the line of continuous Dirichlet (CD) conformal defects as well as the line of continuous Neumann (CN) conformal defects \cite{Defect}.

Finally, we can use a slight modification of the above ansatz to connect two (different) critical semi-infinite lattices and study their interface. This is achieved by considering two types of bulk tensors, namely left bulk tensors $u_L$ and $w_{L}$ for the system on the left and right bulk tensors $u_R$ and $w_R$ for the system on the right, in addition to interface tensors $u_{I}$, $w_{I,L}$ and $w_{I,R}$, see Fig. \ref{fig:defectMERA}(ii). For instance, this ansatz can be used to study fixed-point interfaces between a semi-infinite quantum Ising chain and a semi-infinite quantum XX chain \cite{Prep}.



\begin{thebibliography}{99}

\bibitem{ER} G. Vidal, Phys. Rev. Lett. \textbf{99}, 220405 (2007).

\bibitem{MERA} G. Vidal, Phys. Rev. Lett. \textbf{101}, 110501 (2008). 

\bibitem{MPS} M. Fannes, B. Nachtergaele and R. F. Werner, Comm.
Math. Phys. 144, 3 (1992), pp. 443-490. 
S. \"Ostlund and S. Rommer, Phys. Rev. Lett. 75, 19 (1995), pp. 3537.

\bibitem{Wilson} K.G. Wilson, Rev. Mod. Phys. \textbf{47}, 773 (1975).

\bibitem{DMRG} S. R. White, Phys. Rev. Lett. {\bf 69}, 2863 (1992), 
Phys. Rev. B {\bf 48}, 10345 (1993). 

\bibitem{Topo} M. Aguado, G. Vidal, Phys. Rev. Lett. \textbf{100}, 070404 (2008).
R. Koenig, B. Reichardt, G. Vidal, Phys. Rev. B 79, 195123 (2009).

\bibitem{Free} G. Evenbly, G. Vidal, arXiv:0710.0692v2; 
ibid, arXiv:0801.2449v1.

\bibitem{Transfer} V. Giovannetti, S. Montangero, R. Fazio, Phys. Rev. Lett. \textbf{101}, 180503 (2008).

\bibitem{MERACFT} R. N. C. Pfeifer, G. Evenbly, G. Vidal, Phys. Rev. A \textbf{79}(4), 040301(R) (2009).

\bibitem{Fazio} S. Montangero, M. Rizzi, V. Giovannetti, R. Fazio, Phys. Rev. B \textbf{80}, 113103 (2009). 
V. Giovannetti, S. Montangero, M. Rizzi, R. Fazio, Phys. Rev. A \textbf{79}, 052314 (2009).
 
\bibitem{CFT} P. Di Francesco, P. Mathieu, and D. Senechal, \emph{Conformal Field Theory} (Springer, 1997).

\bibitem{CardyBCFT} J.L. Cardy, Nucl. Phys. B275 200 (1986). J. Cardy, arXiv:hept-th/0411189v2.

\bibitem{Okunishi} K. Okunishi, J. Phys. Soc. Jap. \textbf{76}, 063001 (2007).

\bibitem{MERAalgorithm} G. Evenbly, G. Vidal, Phys. Rev. B \textbf{79}, 144108 (2009). 

\bibitem{chi} We used $\chi=16$ for the MPS, and $\chi=28(22)$ for the lower(upper) indices of a bulk disentangler $u$.

\bibitem{CFTHenkel} M. Henkel, \emph{Conformal Invariance and Critical Phenomena} (Springer, 1999).

\bibitem{g} I. Affleck and A.W.W. Ludwig, Phys. Rev. Lett. 67, 161
(1991)

\bibitem{Holzhey} C. Holzhey. F. Larsen, F. Wilczek, Nucl. Phys. B \textbf{424} 443 (1994).
T.M. Fiola, J. Preskill, A. Strominger, and S. P. Trivedi, Phys. Rev. D \textbf{50} 3987 (1994).

\bibitem{Entropy} G. Vidal, E. Rico, J.I. Latorre, A. Kitaev, Phys. Rev. Lett. \textbf{90} 227902 (2003). V. E. Korepin, Phys. Rev. Lett. 92, 096402 (2004); B. Q.
Jin and V. E. Korepin, J. Stat. Phys. 116, 79 (2004).

\bibitem{Calabrese} P. Calabrese and J. Cardy, J. Stat. Mech. P06002 (2004).

\bibitem{Zhou} H.-Q. Zhou, T. Barthel, J. O. Fjaerestad, U. Schollwoeck, Phys. Rev. A 74, 050305(R) (2006). 

\bibitem{EntropyB} I. Affleck, N. Laflorencie, E. S. Sorensen, J. Phys. A: Math. Theor. \textbf{42} 504009 (2009).

\bibitem{Prep} G. Evenbly et al., \emph{in preparation}.

\bibitem{Defect} M. Oshikawa, I. Affleck, Nucl. Phys. B \textbf{495} 533-582 (1997).
  
\end{thebibliography}
\end{document}